\documentclass[aip,reprint,rsi,numerical,superscriptaddress,prbib]{revtex4-1}
\usepackage{graphicx}
\usepackage{float}
\usepackage{nicefrac}
\newcommand{\Fig}[1]{Fig.~\ref{#1}}

\begin{document}

\author{Clifford W. Hicks} 
\affiliation{Max Planck Institute for Chemical Physics of Solids, N\"{o}thnitzer Stra\ss e 40, Dresden 01187, Germany}
\affiliation{Scottish Universities Physics Alliance (SUPA), School of Physics and Astronomy, University of St.
Andrews, St. Andrews KY16 9SS, United Kingdom}
\author{Mark E. Barber}
\affiliation{Max Planck Institute for Chemical Physics of Solids, N\"{o}thnitzer Stra\ss e 40, Dresden 01187, Germany}
\affiliation{Scottish Universities Physics Alliance (SUPA), School of Physics and Astronomy, University of St.
Andrews, St. Andrews KY16 9SS, United Kingdom}
\author{Stephen D. Edkins}
\affiliation{Scottish Universities Physics Alliance (SUPA), School of Physics and Astronomy, University of St.
Andrews, St. Andrews KY16 9SS, United Kingdom}
\affiliation{Laboratory of Solid State Physics, Department of Physics, Cornell University, Ithaca, NY 14853,
U.S.A.}
\author{Daniel O. Brodsky}
\affiliation{Max Planck Institute for Chemical Physics of Solids, N\"{o}thnitzer Stra\ss e 40, Dresden 01187, Germany}
\affiliation{Scottish Universities Physics Alliance (SUPA), School of Physics and Astronomy, University of St.
Andrews, St. Andrews KY16 9SS, United Kingdom}
\author{Andrew P. Mackenzie}
\affiliation{Max Planck Institute for Chemical Physics of Solids, N\"{o}thnitzer Stra\ss e 40, Dresden 01187, Germany}
\affiliation{Scottish Universities Physics Alliance (SUPA), School of Physics and Astronomy, University of St.
Andrews, St. Andrews KY16 9SS, United Kingdom}

\title{Piezoelectric-based apparatus for strain tuning}

\date{21 May 2014}

\begin{abstract}

We report the design and construction of piezoelectric-based apparatus for applying continuously tuneable
compressive and tensile strains to test samples. It can be used across a wide temperature range, including
cryogenic temperatures. The achievable strain is large, so far up to 0.23\% at cryogenic temperatures. The
apparatus is compact and compatible with a wide variety of experimental probes. In addition, we present a
method for mounting high-aspect-ratio samples in order to achieve high strain homogeneity. {\it The final
version of this article has open access and is available at:} Review of Scientific Instruments {\it vol. 85
article 065003.}

\end{abstract}

\maketitle

\section*{Introduction}

Response to uniaxial pressure can be a powerful probe of the electronic properties of materials. Uniaxial
pressure directly drives anisotropic changes in the nearest-neighbor overlap integrals between atomic sites,
and so will typically drive much larger changes to the electronic structure of materials than equal
hydrostatic pressure. Furthermore, uniaxial pressure is directional, allowing the responses to different
lattice distortions to be compared.

Uniaxial pressure is a well-established technique. To cite just a few results: The superconducting transition
temperature $T_c$ of near-optimally-doped YBa$_2$Cu$_3$O$_{7-\delta}$ increases if the orthorhombicity of the
lattice is artificially reduced by uniaxial pressure.~\cite{Welp92} $T_c$ of
La$_{1.64}$Eu$_{0.2}$Sr$_{0.16}$CuO$_4$ nearly doubles with modest pressure along a $\langle
110 \rangle$ crystal direction, but is less sensitive to $\langle 100 \rangle$ pressures.~\cite{Takeshita04}
The iron pnictide superconductors Ba(Fe,Co)$_2$As$_2$ and BaFe$_2$(As,P)$_2$ are extraordinarily sensitive to
$\langle 110 \rangle$, but not $\langle 100 \rangle$, pressures.~\cite{Kuo12} 

The most common way to apply adjustable pressure to test samples is to clamp the sample between two anvils.
Other methods have also been developed. Adjustable strains have been applied using
bending devices,~\cite{Cao09, Park13} in which bending the substrate changes the sample strain. Another
method is direct attachment of samples to piezoelectric stacks.~\cite{Shayegan03} 

In this article we report the design and construction of a piezoelectric-based strain apparatus in which the
sample is separated from the piezoelectric stacks. The use of piezoelectric stacks gives rapid, precise, {\it
in situ} tunability. The stacks can be made much longer than the sample, so that far larger strains can be
achieved on the sample than on the stacks. Finally, the stacks are arranged in a way that cancels their
thermal contraction, so that the sample can be both tensioned and compressed over a wide temperature range,
including cryogenic temperatures.

Along with precise tunability, high strain homogeneity within the sample was also an important goal of the
present development effort. Strain inhomogeneity has been among the most significant technical difficulties
in uniaxial pressure experiments. Transitions observed under uniaxial pressure have generally broadened,
sometimes severely, as the pressure was increased, an indication of increasing strain
inhomogeneity.~\cite{Torikachvili09, Johnson11, Dix09, Takeshita04} To obtain better strain homogeneity, and
also to allow samples to be tensioned, we discuss the use of epoxy to mount samples with high length-to-width
aspect ratios. We find that high uniaxial pressures, at least 0.4~GPa, can be transmitted through the epoxy.

In the appendices we discuss in some detail elastic deformation of the mounting epoxy, with the aim of
providing a practical guide.

We believe that response to lattice strain is an under-utilised technique. The apparatus and mounting methods
we have developed are compact and reliable, and will allow new experiments across a wide range of materials. 

\section*{Current methods}

We start with a brief discussion of stress and strain. To apply controlled uniaxial stresses to a sample, one
usually compresses a spring, or pressurizes a gas resevoir, which pushes on an anvil that compresses the
sample. In both cases, if the apparatus spring constant is much lower than that of the sample, the
controlled parameter is stress. Conversely, if the apparatus spring constant is much higher than that of the
sample, the controlled parameter is strain: the apparatus applies a displacement to the sample, and the sample
deforms in response to this displacement, ideally independently of its own Young's modulus.

In the linear regime, where stress and strain are linked by a proportionality constant, the distinction
between controlled-stress and controlled-strain apparatus may seem semantic. But there are practical
consequences, the most important of which may be in thermal contraction: in well-designed
controlled-stress apparatus, the spring takes up differential thermal contractions, keeping the force on the
sample essentially constant, but in controlled-strain apparatus one must carefully consider the effects of
differential thermal contraction. Also, if the sample undergoes a structural transition between the mounting
and measurement temperatures, the results from controlled-stress and controlled-strain apparatus will be
qualitatively different.

In a controlled-stress apparatus with anvils to compress the sample, if the sample and anvil faces are in
direct contact then both must be polished flat. In typical uniaxial pressure measurements, the sample strain
is $\sim$0.1\%, corresponding to $\sim$1~$\mu$m of compression over a 1-mm-long sample. Achieving high strain
homogeneity would then require the sample and anvil faces to be smooth, flat, and parallel on a scale well
below this $\sim$1~$\mu$m.  However scientific samples are frequently small, of irregular shape and have
non-ideal mechanical properties for fine polishing; the difficulty in obtaining narrow transitions under
uniaxial pressure suggests that adequate sample polishing is not a trivial task. And even if the sample and
anvil faces match perfectly, frictional locking can introduce strain inhomogeneity: the end faces of the
sample are locked to the anvils, while the center attempts to expand following its own Poisson's
ratio.~\cite{Wei09}

Strain homogeneity can be improved by using samples with higher aspect ratios (length over width): the effects
of irregularities (that do not generate bending moments) at the sample-anvil interface decay towards the
sample center, and the sample's Poisson's ratio dominates its transverse strain. In
Ref.~\onlinecite{Bourdarot11}, an aspect ratio of 2:1 --- high for uniaxial pressure experiments on
correlated-electron materials --- was used to improve the strain homogeneity. Also, gold and cadmium films
were inserted at the sample-anvil interface, to reduce frictional locking and smooth out defects.

In most uniaxial pressure apparatus the pressure is set at room temperature, by turning a bolt. {\it In situ}
adjustability has been achieved in low-temperature apparatus by using helium-filled bellows to apply the
force.~\cite{Welp92, Pfleiderer97, Dix09} 

Direct attachment of samples to piezoelectric stacks offers {\it in situ} adjustability, in a much
simpler and more compact apparatus. This technique was introduced in Ref.~\onlinecite{Shayegan03} for strain tuning
of semiconductors, and extended to correlated electron materials in Ref.~\onlinecite{Chu12}. However there are two
significant limitations of the sample-on-stack technique: limited range, and large differential thermal
contraction. In our apparatus we used lead zirconium titanate (PZT) stacks,~\cite{piezoSource} the most common
composition, and the catalog indicates that at room temperature, within the manufacturer's recommended voltage
limits of $-30$ and $+150$~V, the total range of strain on the stacks is $\sim$0.15\%. This is small: when we
tested our apparatus with a sample of Sr$_2$RuO$_4$, we found that the samples snapped under $\approx$0.25\%
tension, and could withstand at least the same amount of compression, meaning that the sample itself permitted
a strain range of at least 0.5\%. Microscopic VO$_2$ rods have been found to withstand up to 2.5\%
strain,~\cite{Cao09} and, for an extreme case, it is calculated that defect-free silicon nitride could
withstand tensile strains of up to $\sim$25\%.~\cite{Ogata01}

Furthermore, the response of the piezoelectric stacks falls as the temperature is reduced. At
$\sim$1~K, we found the response per volt of our stacks (measured using a strain gauge) to be $1/6$ that at
room temperature.~\cite{strainGaugeSource, responseFootnote} This reduced response can be partially offset by
the larger voltages that can be applied at cryogenic temperatures: a 0.04\% strain range (-0.02\% to +0.02\%)
was obtained with voltages between $-300$ and $+300$~V (on a different piezoelectric stack model from the same
manufacturer),~\cite{Shayegan03} while we achieved a 0.05\% range over $-170$ to $+420$~V. This is still at
least an order of magnitude less than the strain range that typical samples can withstand.

Large differential thermal contraction is a challenge because PZT {\it lengthens} along its poling direction
as it is cooled, by $\sim$0.1\% between room temperature and 4~K.~\cite{Simpson87, PIdatasheet,
expansionFootnote} Very few materials contract by less than 0.1\% over this range; 0.2--0.3\% is more typical.
Therefore (and in the absence of any plastic deformation of the mounting epoxy~\cite{creepFootnote})
differential thermal contraction will strain typical samples by an amount well beyond the range of the stacks,
making it impossible to tune the strain through zero.  Overall, the sample-on-stack technique is best suited
for measuring the linear response to small strain perturbations,~\cite{Kuo13} in circumstances where a
significant nonlinear contribution is not expected. 

\section*{The uniaxial strain apparatus}

\begin{figure}
\includegraphics[width=3.25in]{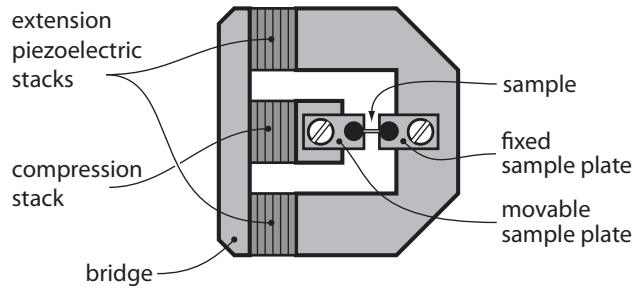}
\caption{\label{overview}A schematic overview of the strain apparatus}
\end{figure}

A schematic overview of our apparatus is shown in \Fig{overview}. The sample is firmly affixed with epoxy
across a gap between two plates, one movable and the other fixed. The position of the movable plate is
actuated by three piezoelectric stacks, which are joined by a bridge. A positive voltage applied to the
central stack extends the stack and compresses the sample, while a positive voltage on the outer two stacks
pushes the bridge outwards and tensions the sample. All the stacks have equal lengths, so in principle their
thermal expansion does not strain the sample. 

Because the stacks are much longer than the sample, large sample strains are achievable. The sample strain is
$(L_{\mathrm{st}}/L_{\mathrm{sa}}) \times ( \varepsilon_{\mathrm{outer}} - \varepsilon_{\mathrm{central}} )$,
where $L_{\mathrm{st}}$ is the length of the stacks, $L_{\mathrm{sa}}$ is the strained length of the sample,
and $\varepsilon_{\mathrm{outer}}$ and $\varepsilon_{\mathrm{central}}$ are the strains on the outer and
central stacks. (The ``strained length'' of the sample is the length over which strain is applied: as will be
described, the sample is mounted with epoxy in a way that strain is not applied to the end portions of the
sample.) In our first apparatus, $L_{\mathrm{st}}$ was 4~mm, and $L_{\mathrm{sa}}$ typically around 1~mm; we
achieved sample strains below 4~K of up to 0.23\%.~\cite{Hicks14}

Our terminology requires some discussion. The apparatus is accurately described as a uniaxial strain apparatus, but the
applied strain is not strictly uniaxial, and the control over strain is not perfectly rigid. The sample will
have nonzero Poisson's ratios, so strain applied along its length will induce strains along its width and
thickness. However the {\it stress} within the sample is strictly uniaxial, and the apparatus offers
independent control of the strain along only a single axis, so its description as uniaxial is appropriate.
The control over the strain is not perfectly rigid because, although the apparatus itself is several times
stiffer than typical samples,~\cite{stiffnessFootnote} the epoxy used to mount the sample deforms, taking up
some of the applied displacement. For the samples this apparatus was designed to accept, high-Young's-modulus
crystals with cross-sectional areas $\sim$0.01~mm$^2$, the epoxy spring constant remains higher than the
sample spring constant (as detailed in the appendices), but not so high that epoxy deformation can be ignored
in determining the sample strain. The description of this as a controlled-strain apparatus is appropriate
because samples could in principle be mounted more rigidly, and it is important to retain a clear distinction
with controlled-stress apparatus, in which there must be a well-defined spring of some form with a low spring
constant.

\begin{figure}
\includegraphics[width=3.25in]{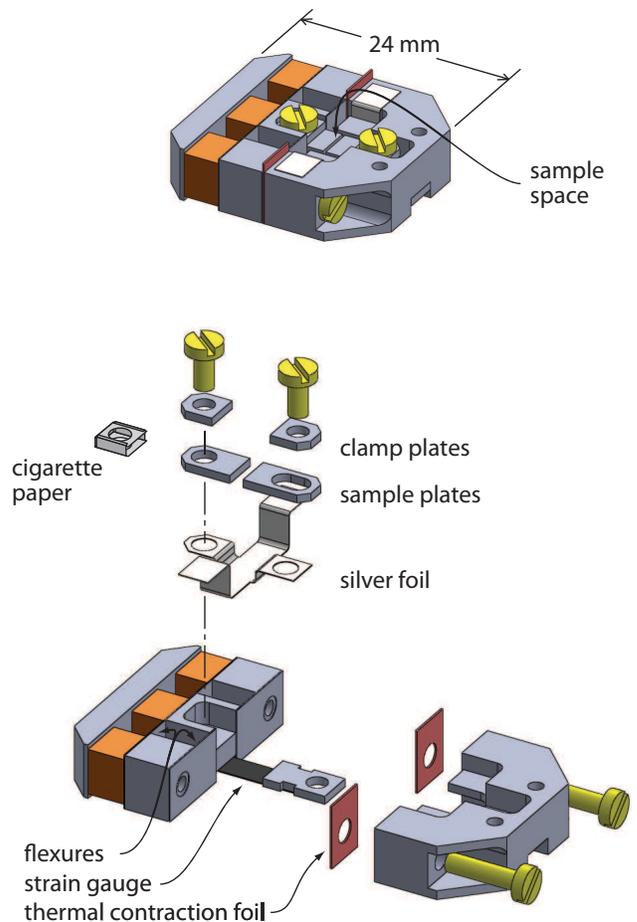}
\caption{\label{explodedView}Our strain apparatus.}
\end{figure}

\Fig{explodedView} shows the complete apparatus; we now describe some of the details.

The flexures present a low spring constant for longitudinal motion, and a much higher spring constant for 
twisting or transverse motions. They are intended to protect the stacks from inadvertent transverse forces,
for example during the sample mounting process, and to reduce unwanted bending from loads not centered on the
stacks.

Our first device was constructed out of titanium, chosen because its thermal contraction is similar to the
transverse thermal contraction of the stacks.~\cite{Simpson87, PIdatasheet} This thermal contraction is lower
than most materials, however, so differential thermal contraction would place most samples under tension.
Copper foils (the ``thermal contraction foils'' in the figure) were incorporated to increase the device's
thermal contraction. The screws holding the apparatus together are brass, which contracts more than titanium,
and so secure the apparatus more tightly as it cools.

A strain gauge~\cite{strainGaugeSource} was incorporated to measure the displacement applied to the sample,
from which the sample strain could be calculated. The piezoelectric stacks in our first apparatus are
hysteretic, particularly at large voltages, so a position sensor is necessary. The gauge is mounted across a
6-mm-wide gap beneath the sample; the samples are far too small for a gauge to be affixed directly to them.
To stiffen the gauge and reduce deformation during handling, it was first epoxied to a piece of cigarette
paper. The gauge and cigarette paper combination was then mounted under tension, so that it would remain flat
even if the sample was strongly compressed. 

The strain gauge was not a perfect sensor in that its resistance had a small temperature dependence (over our
initial measurement range of 0.5--3~K), and shifted slightly but noticably between cool-downs. However these
effects could be treated during data analysis, and the gauge provided a non-hysteretic measure of the sample
strain within each cool-down.

The silver foil is intended to reduce the thermal time constant between the sample and a temperature sensor
mounted on a free tab of the foil. Cigarette paper can be used to electrically isolate one or both of the
sample plates, if desired.
\begin{figure}[ptb]
\includegraphics[width=3.25in]{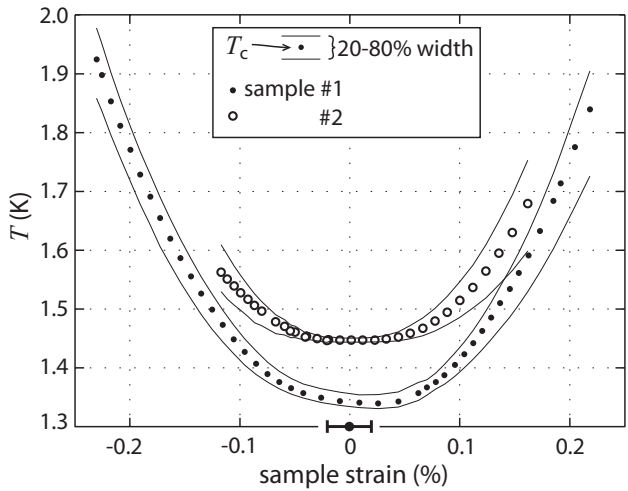}
\caption{\label{214results} $T_c$ of Sr$_2$RuO$_4$ under strain applied along a $\langle 100 \rangle$ crystal
direction. The points are where the magnetic susceptibility reached 50\% of its normal-state value, and the
lines 20 and 80\%, giving a measure of the transition widths.  The error bar on the $x$-axis is the error in
locating zero strain. Reprinted from Ref.~\onlinecite{Hicks14}.} 
\end{figure}

The stacks can be operated together to achieve smooth strain ramps. For example, to sweep the strain from
strong compression to strong tension, the voltages on the (compression, tension) stacks might be ramped from
(300, 0) to (150, 150) to (0, 300)~V, thus avoiding discontinuity in operation across zero strain.

The size limit of samples that this apparatus can accept is currently unclear. Force applied to the
sample places at least one of the stacks under tension, but piezoelectric stacks are sintered powders,
not meant to withstand high tensile stress. In our first experiment, the applied force never exceeded 5~N.
The stacks can likely withstand considerably larger tensile forces than that. 

In \Fig{214results}, we show data collected with this apparatus: the superconducting transition temperature
$T_c$ against strain (applied along a $\langle 100 \rangle$ crystal direction) for two single crystals of the
unconventional superconductor Sr$_2$RuO$_4$. The sample cross-sections were 110$\times$30 and
170$\times$60~$\mu$m. The Young's modulus of Sr$_2$RuO$_4$ is 182~GPa,~\cite{Paglione02} so at the highest 
strains the stress in the sample was about 0.4~GPa. $T_c$ of Sr$_2$RuO$_4$ increases strongly both when it is
tensioned and compressed. The data in the figure illustrate the capabilities of the apparatus: the rapid
and precise tunability allowed a high density of data points, and the curves are smooth. The scientific
results of this experiment are discussed in Ref.~\onlinecite{Hicks14}.

\section*{Sample mounting}

The apparatus was designed to accept, initially, samples with cross sections of $\sim$200$\times$50~$\mu$m.
The samples needed to be epoxied into place, both so that they could be tensioned, and to reliably
transmit the micron-scale displacements generated by the piezoelectric stacks. But mounting with epoxy gives
other advantages. One is that the epoxy conforms to the sample, so precision polishing of the sample faces is
not necessary. The samples do need to be cut to have an approximately constant cross-section, but the demands
on precision here are not severe. Another advantage is that the sample ends cannot pivot, which allows higher
length-to-width aspect ratios before the sample buckles under compression. Finally, if the epoxy has
relatively low elastic moduli, it forms a deformable interface layer that reduces stress concentration in the
sample, reducing the risk of sample fracture.

\begin{figure}[ptb]
\includegraphics[width=3.25in]{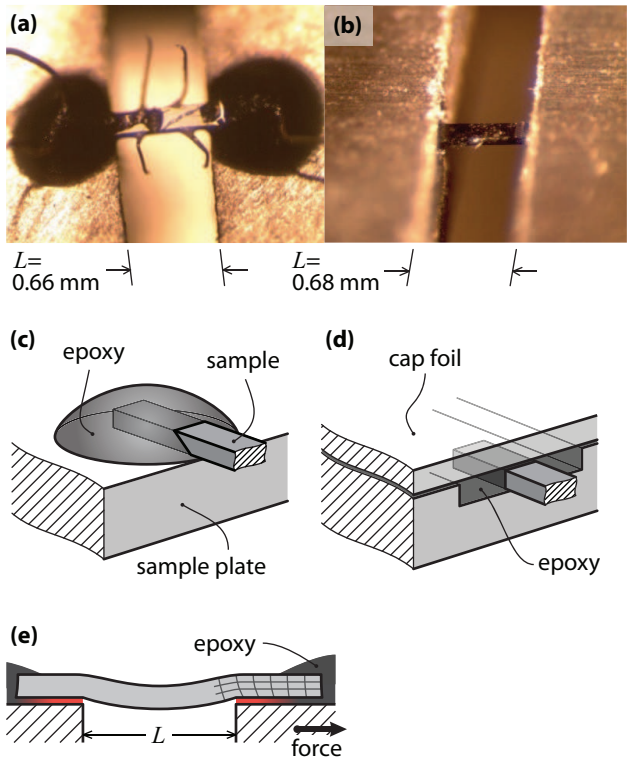}
\caption{\label{mountedSamples}(a) and (b): samples mounted across the gap between the two titanium sample
plates, by two different methods, and (c) and (d): schematics of the structures in (a) and (b). The epoxy is
Stycast 2850FT. In panel (a), the wires attached to the sample were used for resistance measurements prior to
mounting in the strain device. (e) Illustration of how a sample mounted as in panel (a) deforms when
tensioned; most of the load on the sample is transferred through the portions of epoxy shaded red.}
\end{figure}

We used Stycast\textsuperscript{\textregistered} 2850FT. Early samples were mounted as shown in
\Fig{mountedSamples}(a), with droplets of epoxy securing the ends, and no further construction. While simple,
the disadvantage of this method is its asymmetry: the sample is secured more firmly through its lower than its
upper surface. A calculation presented in Appendix~A shows that it is the leading $\sim$0.1~mm of the epoxy,
shaded red in panel (e) of the figure, that transfers most of the applied force between the sample plate and
sample. Due to the asymmetry, when the sample is strained it also bends, downward when tensioned and upward
when compressed. The bending introduces a strain gradient in the sample, which, as shown in \Fig{bending}, in
Appendix B, can be substantial.

Later samples were therefore mounted as shown in \Fig{mountedSamples}(b): with a rigid cap foil over the
sample, so that the sample is secured through both its lower and upper surfaces. In
\Fig{214results}, sample \#1 was mounted in this way, and sample \#2 as in panel (a). The superconducting
transition of sample \#2 broadened considerably more under strain than that of sample \#1, indicating greater
strain inhomogeneity.

As noted in the introduction, the epoxy mounts were sufficient to transmit sample pressures of at least
0.4~GPa. We also tested the epoxy at room temperature, by tensioning samples mounted as in panel (a) until
fracture. We tested two samples with Epotek\textsuperscript{\textregistered} H20E epoxy and one with Stycast
2850FT.  The samples were 70--120~$\mu$m wide and 30--100~$\mu$m thick. In all three cases, the samples snapped
at tensions of $\sim$0.25\%. Fracture occured towards the middle of the sample: it was the sample, not the
epoxy, that failed. For larger samples with a lower surface-area-to-volume ratio, the stress in the epoxy will
be higher and eventually the strength of the epoxy will become the limiting factor, but it is clear that
there is a practical range of parameters where high sample pressures can be achieved.

We worked with samples with length-to-width aspect ratios $L/w$ between 3.5 and 7. ($L$ here and in the
appendices refers to the exposed length of the sample, ignoring the end portions that are embedded in epoxy.)
In retrospect, seven was more than necessary. As discussed in Appendix B, if the epoxy has low elastic moduli
and the epoxy layers are sufficiently thick (at least $\sim \nicefrac[]{1}{4}$ the sample thickness), the
strain within most of the exposed portion of the sample is highly homogeneous, with significant inhomogeneity
(apart from any bending-induced gradients) only very near to the sample mounts (Appendix B, Table~I). 

There are also advantages in working with samples that are thin plates, with $w/t$ ($t$ the sample thickness)
significantly greater than one. (For Sr$_2$RuO$_4$, a layered compound, this was a natural geometry.) The
surface-area-to-volume ratio is increased, reducing stress within the epoxy, and bending-induced strain
variation is reduced. If both $L/w$ and $w/t$ are significantly greater than one, however, $L/t$ can become
quite large; the highest $L/t$ in our Sr$_2$RuO$_4$ experiment was 25. The Euler formula for the buckling load
on a thin beam with both ends unable to pivot is
\[
F = \frac{4 \pi^2 E I}{L^2},
\]
where $E$ is the Young's modulus and $I$ is the area moment of inertia.~\cite{Euler} $I$ for a thin rectangular plate is
$t^3 w/12$, and the longitudinal strain is $\varepsilon = F/Ewt$. Substituting, the critical aspect ratio
$L/t$, above which the plate buckles, is
\[
\frac{L}{t} = \frac{\pi}{\sqrt{3 \varepsilon}}.
\]
For $\varepsilon = 0.25$\%, the sample is expected to buckle for $L/t>36$.

\section*{Conclusion}

We have presented a design for compact, piezoelectric-based apparatus that can apply large strains to test
samples, even at cryogenic temperatures. The apparatus can apply both compressive and tensile strains, a
useful technological advance. We have also discussed and analysed a method for obtaining high strain
homogeneity within the sample, whether using this or another distortion apparatus.

We anticipate that apparatus and methods similar to those presented here will be widely applicable.
Strain-tuning is conceptually a very simple technique, and we believe that much can be learned across many
systems from basic measurements such as resistivity and magnetic susceptibility as a function of strain. This
apparatus also leaves the upper surface of the sample exposed, allowing access for spectroscopic and
scattering probes. In summary, we hope that the methods that we have presented will make strain-tuning a more
practical, widespread and precise technique.

\begin{acknowledgements}

The authors thank Jan Bruin, Ian Fisher, Andrew Huxley and Edward Yelland for valuable discussions. They thank
the UK Engineering and Physical Sciences Research Council and the Max Planck Society for financial support.

\end{acknowledgements}

\appendix

\section{Analytic analysis of the sample mounts.}

\begin{figure}
\includegraphics[width=3.25in]{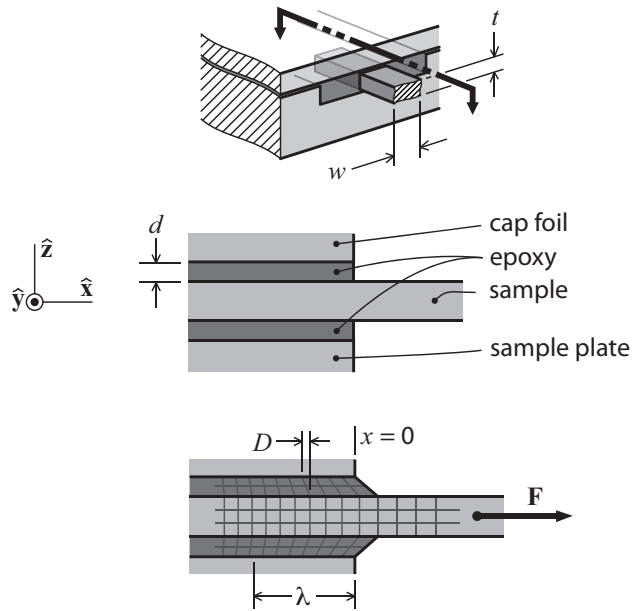}
\caption{\label{deformationModel} A model for estimating the load transfer length $\lambda$, here for the
mounting method in panel (d) of \Fig{mountedSamples}. When a force $\mathbf{F}$ is applied to the sample, the
load is transferred to the sample plate and cap foil -- both taken to be perfectly rigid -- over a length
scale $\lambda$. $D(x)$ is the $x$-dependent displacement of the sample.}
\end{figure}

In Appendix A, we estimate analytically the load transfer length $\lambda$, the length over which the
applied force is transferred between the sample plates and sample. The displacement applied by the
piezoelectric stacks, and measured by the strain gauge, will be distributed over a length $L+2\lambda$, so
knowledge of $\lambda$ is needed to estimate the sample strain. We also discuss the stress within the epoxy.
The parameters for our model are illustrated in \Fig{deformationModel}. We make the following simplifications:
(1) The sample width $w$ is sufficiently larger than its thickness $t$ that bonding on the sides of the sample
is not important. (2) The sample plate and cap foil are perfectly rigid. (3) Shears within the sample are
neglected: the strain within the sample, $\varepsilon_{xx}(x)$, is constant in both $y$ and $z$.  These latter
two assumptions amount to supposing that the epoxy has much lower elastic constants than the sample, sample
plate, and cap foil.

Within this model, the force within the sample at position $x$ is $F(x) = E w t \varepsilon_{xx}$, where $E$
is the Young's modulus of the sample.~\cite{elasticModulusFootnote} $F$ varies with $x$ following:
\[
\frac{dF}{dx} = n w \sigma(x) \approx n w C_{66,\mathrm{e}}\frac{D(x)}{d},
\]
where $\sigma$ is the shear stress across the interface between the sample and epoxy, $C_{66, \mathrm{e}}$ the
shear elastic constant of the epoxy, $d$ the epoxy thickness, and $D(x)$ the displacement of the sample at
position $x$ from its unloaded position. $n$=1 if the sample is bonded on its lower side only, and 2 if on
both the top and bottom. $\varepsilon_{xx}$ and $D$ are related by $\varepsilon_{xx}=dD/dx$, so a differential
equation for $D$ can be readily obtained and solved. Its solution is $D$ decaying exponentially over a length
scale
\[
\lambda = \sqrt{\frac{Etd}{nC_{66, \mathrm{e}}}}.
\]

The elastic properties of Stycast~2850FT appear not to have been measured precisely at cryogenic temperatures.
In a technical study for spacecraft applications, its Young's modulus was found to increase gradually as the
temperature was reduced, but appeared to level off below $\sim$160~K.~\cite{JPLreport} At 150~K, it was
determined to be $11\nicefrac[]{1}{2}$~GPa when Catalyst~24LV was used, and 16~GPa when Catalyst~9 was used.
(We used Catalyst 23~LV.) The Young's modulus of Stycast~1266, an unfilled version of 2850FT, has been
measured at 197~K, 77~K, and a few temperatures between 77 and 2.2~K;~\cite{Hashimoto80} it was found to be
$\approx$4.5~GPa for temperatures 77~K and below. If $E$ of Stycast 2850FT behaves similarly, it may rise
slightly from its 150~K value as the temperature is reduced further, before leveling off.

The shear modulus of an isotropic material is $C_{66} = E/2(1+\nu)$, where $\nu$ is Poisson's ratio. We take
$E \sim 15$~GPa and $\nu \sim 0.3$, yielding $C_{66, \mathrm{e}} \sim 6$~GPa for the Stycast.

Sr$_2$RuO$_4$ is a relatively stiff material, with $E = 182$~GPa.~\cite{Paglione02} Taking typical values
$t=50$~$\mu$m, $d=10$~$\mu$m and $n=2$ yields $\lambda \approx 90$~$\mu$m: it is the leading $\sim$0.1~mm of
epoxy that transfers the applied displacement to the sample.

The shear strain within the epoxy, $\varepsilon_{xy, \mathrm{e}}$, will be maximal at the edge of the
sample plate, $x=0$, where it is:
\begin{equation}
\varepsilon_{xy,\mathrm{e}}(0) = \frac{\varepsilon_\mathrm{app} \lambda}{d} = \varepsilon_\mathrm{app}
\sqrt{\frac{Et}{ndC_{66,\mathrm{e}}}},
\end{equation}
where $\varepsilon_\mathrm{app}$ is the sample strain beyond the end of the epoxy. For the above parameters,
$\varepsilon_{xy,\mathrm{e}}(0)$ comes to 1.3\% for $\varepsilon_{\mathrm{app}}=0.2$\%.

The data sheet for Stycast 2850FT indicates a tensile strength of $\sim$50~MPa (at room
temperature).~\cite{stycastDatasheet} With $\varepsilon_{\mathrm{app}}=0.2$\%, the shear stress 
in our sample mounts, using the above parameters, is $C_{66, \mathrm{e}} \times \varepsilon_{xy, \mathrm{e}} =
80$~MPa at $x=0$. We may therefore have been close to the yield strength of the epoxy.
The measurements on Stycast~1266 however indicate a fracture strain of $\sim$4\% at low
temperatures,~\cite{Hashimoto80} and if Stycast 2850FT performs similarly then our mounts had a comfortable
margin of safety. Our measurements showed almost no hysteresis against strain, and no abrupt changes in
behavior at high strains, indicating that the epoxy did not fracture or de-bond. 

If failure of the epoxy becomes a significant limitation in future measurements, Eq.~A1 indicates the steps to
take: The sample should be bonded from both sides (so that $n=2$). The sample should be made thin, and the
epoxy layer somewhat thick.  The shear stress at the interface is $\propto$$\sqrt{C_{66,\mathrm{e}}}$, so a
good choice of epoxy appears to be one with low elastic constants, high bonding strength and high yield
strain.
\\

\section{Finite-element analysis}

Here we present the results of finite element simulation of a few representative cases. We discuss the load
transfer length $\lambda$, strain homogeneity and sample bending.

\begin{figure}
\includegraphics[width=3.25in]{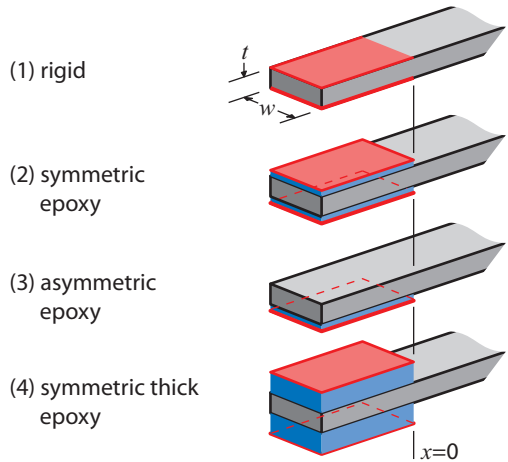}
\caption{\label{mountModels} Models used for finite element calculation. Red indicates the fixed faces, and
blue layers are epoxy.}
\end{figure}

We study four models for the sample mounts, illustrated in \Fig{mountModels}. They are: (1) ``Rigid:''
the sample is secured perfectly rigidly on its top and bottom surfaces.  (2) ``Symmetric epoxy:'' the
sample is bonded on both its top and bottom faces through thin layers of low-elastic-modulus epoxy to
perfectly rigid surfaces (the sample plate and cap foil). (3) ``Asymmetric epoxy:'' only the lower
surface is bonded, again with relatively soft epoxy.  (4) ``Symmetric thick epoxy:'' same as \#2, but
with thicker epoxy layers. Models \#2 and 3 are close to our actual conditions, in which the samples
were $\sim$50~$\mu$m thick, and the epoxy 10--20~$\mu$m thick. Models \#1 and 4 are included for
comparison.

There are a few parameters to specify. For the sake of generality, we take both the epoxy and sample to be
isotropic, with a Poisson's ratio of 0.3. Young's modulus for the epoxy is set to
\nicefrac[]{1}{10} that of the sample. The thickness of the epoxy layers is set to $0.25t$ for the thin
layers and $t$ for the thick layers. $w$ is set to $4t$, and $L$ to $6w$.

The calculations were done using a rectilinear mesh, with 15 or 16 elements spanning each of the sample
thickness, sample width and epoxy thickness. The portions of the sample embedded in the epoxy were in all
cases made much longer than the load transfer length $\lambda$. Differential thermal contractions are
neglected.
\begin{figure}
\includegraphics[width=3.25in]{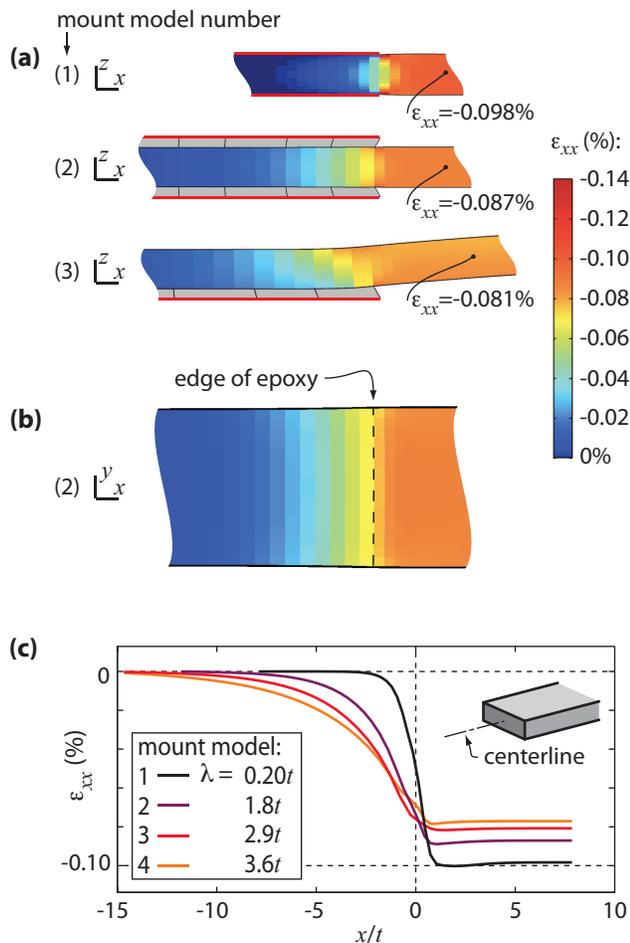}
\caption{\label{deformedSamples}Strain $\varepsilon_{xx}$ for samples mounted as in the models of
\Fig{mountModels}. In all cases, the movable sample plate was moved inward by 0.1\% of $L$.
 (a) $\varepsilon_{xx}$ in the $xz$ center planes of samples mounted as in models (1) through (3). The red
lines indicate the fixed faces. Deformations are exaggerated by a factor of 100. (b) $\varepsilon_{xx}$ in the
$xy$ center plane for mount model (2). (c) $\varepsilon_{xx}$ along the centerline for all the mount models.
The inset shows results for $\lambda$, determined as described in the text.}
\end{figure}

Fig.~\ref{deformedSamples} shows some results for the strain $\varepsilon_{xx}$. In all cases, the movable
sample plate was moved inward by 0.1\% of $L$, but because $\lambda>0$, the actual sample strain in the gap is
somewhat less than 0.1\%.  In panel (c), we report $\lambda$ for each calculation, determined such that to
achieve an applied strain $\varepsilon_{\mathrm{app}}$ in the gap, the movable sample plate should be moved a
distance $\varepsilon_{\mathrm{app}}(L + 2\lambda)$. 

$\lambda$ depends on parameters such as the epoxy thickness that, in practice, can be difficult to control 
accurately, particularly for small samples. One disadvantage in mounting samples with low-elastic-modulus
epoxy is that greater absolute uncertainty in $\lambda$ means greater uncertainty in the sample strain.
However, the results also show that stress concentration within the sample is reduced.

We next discuss strain homogeneity. Provided the sample does not bend (that is, the mounts are symmetric),
strain inhomogeneity will decay exponentially towards the sample center; measurements should be configured to
be sensitive mainly to the sample center. A guide on how much of the ends of the sample (in addition to the
portions embedded in the epoxy) to exclude is given in Table~I. The criterion is that at some location in the
sample cross-section, the strain $\varepsilon_{xx}$ differs from $\varepsilon_{xx}$ at the sample centre (at
$x=L/2$) by more than a given percentage. For example, using mount model \#2, to obtain less than 5\% strain
inhomogeneity over the entire measured region only the outermost portions of length $0.2w$ need to be excluded
from measurement. In other words, by using suitable sample mounts high strain homogeneity can be
obtained within almost the entire exposed portion of the sample.
\begin{table}[H]
\caption{Lengths of the end portions of sample to exclude from measurement, to obtain a given level of strain
homogeneity. Further explanation is given in the text.}
\begin{tabular*}{\columnwidth}{@{\extracolsep{\fill}}cccccc}
\hline\hline
\% inhomogeneity & Mount model \#1 & \#2 & \#4\\
\hline
5\% & $0.4w$ & $0.2w$ & $0.1w$ \\
1\% & $0.8w$ & $0.6w$ & $0.4w$ \\
\end{tabular*}
\end{table}

\begin{figure}[h]
\includegraphics[width=3.25in]{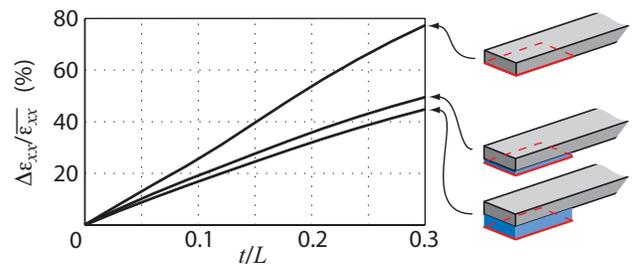}
\caption{\label{bending} Bending-induced strain variation in the middle of the sample (at $x=L/2$) against
$t/L$. $\Delta \varepsilon_{xx} / \overline{\varepsilon_{xx}}$ is the difference between strain at the top and bottom
surfaces, divided by the average strain through the thickness of the sample.  The three cases are: (1) the
bottom surface of the sample is held fixed; (2) and (3) the lower surface is mounted through a layer of
low-elastic-modulus epoxy, with thicknesses $0.25t$ and $t$.}
\end{figure}

If the sample does bend, a strain gradient is introduced into the sample. Let $\Delta \varepsilon_{xx}$ be the
difference between the strains at the upper and lower surfaces of the sample, and
$\overline{\varepsilon_{xx}}$ be the average strain through the thickness of the crystal. Ideally the ratio $\Delta
\varepsilon_{xx} / \overline{\varepsilon_{xx}}$ should be as small as possible. But it may also be desirable to bond the
sample only by its lower surface, for unfettered access to its upper surface, and even if symmetric
sample mounts are constructed imperfection in assembly will lead to residual asymmetry. So it is useful to
know how large $\Delta \varepsilon_{xx}$ might be. In \Fig{bending}, we show calculations of $\Delta
\varepsilon_{xx}/\overline{\varepsilon_{xx}}$ against sample thickness for samples bonded from below only.
Unsurprisingly, $\Delta \varepsilon_{xx}$ is larger for thicker samples.  However the magnitude is
noteworthy: $L/t=20$, for example, is a large aspect ratio not far below the buckling
limit, but $\Delta \varepsilon_{xx} / \overline{\varepsilon_{xx}}$ could still be up to 10\%.  Although slightly more
difficult to implement, symmetric mounting as illustrated in panel (d) of \Fig{mountedSamples} offers a clear
advantage.


\begin{thebibliography}{99}

\bibitem{Welp92}U. Welp, M. Grimsditch, S. Fleshler, W. Nessler, J. Downey and G.W. Crabtree, {\it Phys. Rev.
Lett.} {\bf 69}, 2130 (1992).

\bibitem{Takeshita04}N. Takeshita, T. Sasagawa, T. Sugioka, Y. Tokura and H. Takagi, {\it J. Phys. Soc.
Japan} {\bf 73}, 1123 (2004).

\bibitem{Kuo12}H.-H. Kuo, J.G. Analytis, J.-H. Chu, R.M. Fernandes, J. Schmalian and I.R. Fisher, {\it Phys.
Rev. B} {\bf 86}, 134507 (2012).

\bibitem{Cao09}J. Cao, E. Ertekin, V. Srinivasan, W. Fan, S. Huang, H. Zheng, J.W.L. Yim, D.R. Khanal, D.F.
Ogletree, J.C. Grossman, and J. Wu, {\it Nature Nanotechnology} {\bf 4}, 732 (2009).

\bibitem{Park13}J.H. Park, J.M. Coy, T.S. Kasirga, C.-M. Huang, Z.-Y. Fei, S. Hunter, and D.H. Cobden, {\it
Nature} {\bf 500}, 431 (2013).

\bibitem{Shayegan03}M. Shayegan, K. Karrai, Y. P. Shkolnikov, K. Vakili, E. P. De Poortere and S. Manus, {\it
Appl. Phys. Lett.} {\bf 83}, 5235 (2003).

\bibitem{Torikachvili09}M.S. Torikachvili, S.L. Bud'ko, N. Ni, P.C. Canfield and S.T. Hannahs, {\it Phys.
Rev. B} {\bf 80}, 014521 (2009).

\bibitem{Johnson11}S.D. Johnson, R.J. Zieve and J.C. Cooley, {\it Phys. Rev. B} {\bf 83}, 144510 (2011).

\bibitem{Dix09}O.M. Dix, A.G. Swartz, R.J. Zieve, J. Cooley, T. R. Sayles and M. B. Maple, {\it Phys. Rev.
Lett.} {\bf 102}, 197001 (2009).

\bibitem{Wei09}X.X. Wei and K.T. Chau, {\it Int. J. Solids and Struct.} {\bf 46}, 1953 (2009).

\bibitem{Bourdarot11}F. Bourdarot, N. Martin, S. Raymond, L.-P. Regnault, D. Aoki, V. Taufour and J.
Flouquet, {\it Phys. Rev. B} {\bf 84}, 184430 (2011).

\bibitem{Pfleiderer97}C. Pfleiderer, E. Bedin and B. Salce, {\it Rev. Sci. Instrum.} {\bf 68}, 3120 (1997).

\bibitem{Chu12}J.-H. Chu, H.-H. Kuo, J.G. Analytis and I.R. Fisher, {\it Science} {\bf 337}, 710 (2012).

\bibitem{piezoSource}Pch 150/$5 \times 5$/2, Piezomechanik GmbH.

\bibitem{Ogata01}S. Ogata, N. Hirosaki, C. Kocer and H. Kitagawa, {\it Phys. Rev. B} {\bf 64}, 172102 (2001).

\bibitem{strainGaugeSource}Vishay Micro-Measurements EK-06-250PD-10C/DP. We took the gauge constant, the rate
of change of gauge resistance against variation in the gauge's length, to be temperature-independent.  Vishay
Micro-Measurements Tech Note TN-504-1 (``Strain gauge thermal output and gauge factor variation with
temperature'') indicates that the gauge constant for the Karma Alloy used in our gauges increases, with a
linear temperature dependence, by 1.0\% from 24$^\circ$C to -73$^\circ$C. Extrapolating to 0~K, the gauge
constant would be $\sim$3\% larger than at room temperature. 

\bibitem{responseFootnote}The response rates of the stacks were determined below 80 (200)~V at room temperature
(4 K), where the response was nearly linear with applied voltage.

\bibitem{Simpson87}A.M. Simpson, and W. Wolfs, {\it Rev. Sci. Inst.} {\bf 58}, 2193 (1987).

\bibitem{PIdatasheet}Physik Instrumente GmbH, ``Piezo Material Data.''

\bibitem{expansionFootnote}The Physik Instrument piezo materials datasheet indicates a coefficient of thermal
expansion for various PZT formulations of $-4$ to $-6 \cdot 10^{-6}$/K, along the poling
direction. The thermal contraction of most materials is much diminished below $\sim$77~K, so multiplying this
coefficient by a $\sim$200~K temperature range yields an expansion of 0.08 to 0.12\% from room to
cryogenic temperatures.

\bibitem{creepFootnote}Both Refs.~\onlinecite{Shayegan03} and ~\onlinecite{Chu12} report that much less strain is
transmitted from the stack to the sample at higher temperatures, 300~K for the former and above $\sim$100~K
for the latter, than at low temperatures, suggesting significant plastic deformation of the epoxy at higher
temperatures. This may hinder measurements at higher temperatures, but could have the benefit of relieving
thermal strains.

\bibitem{Kuo13}H.-H. Kuo, M.C. Shapiro, S.C. Riggs and I.R. Fisher, {\it Phys. Rev. B} {\bf 88}, 085113
(2013).

\bibitem{Hicks14}C.W. Hicks, D.O. Brodsky, E.A. Yelland, A.S. Gibbs, J.A.N. Bruin, K. Nishimura, S. Yonezawa,
Y. Maeno and A.P. Mackenzie, {\it Science} {\bf 344}, 283 (2014).

\bibitem{stiffnessFootnote}With a Young's modulus of $\sim$200~GPa, the spring constant for straining the sample
lengthwise will be $Ewt/L \sim 2 \cdot 10^6$~N/m, taking $wt \sim 0.01$~mm$^2$ and $L \sim 1$~mm. The least
stiff part of the apparatus is the bridge, which can be viewed approximately as two S-bending cantilevers
6~mm wide, 2.5~mm thick and 9~mm long, yielding a spring constant of $14 \cdot 10^6$~N/m.

\bibitem{Paglione02}J.P. Paglione, C. Lupien, W.A. MacFarlane, J.M. Perz, L. Taillefer, Z.Q. Mao and Y. Maeno, 
{\it Phys. Rev. B} {\bf 65} 220506 (2002).

\bibitem{Euler}M. Euler, {\it Memoires de l'academie des sciences de Berlin} {\bf 13}, 252 (1759).

\bibitem{elasticModulusFootnote} The Young's modulus for loads along $\mathbf{x}$ is
$E=C_{11}-C_{12}^2/(C_{11}+C_{12})-C_{13}^2/(C_{11}+C_{13})$. It applies if the sample is free to expand and contract,
following the Poisson's ratios, along $\mathbf{y}$ and $\mathbf{z}$. If the sample is a thin plate it can
probably contract along $\mathbf{z}$, but not along $\mathbf{y}$. In this case, the elastic constant
$C_{11}-C_{13}^2/(C_{11}+C_{13})$ should be used instead of $E$. For realistic materials, this is not very
different from $E$.

\bibitem{JPLreport}C.E. Ojeda, E.J. Oakes, J.R. Hill, D. Aldi and G.A. Forsberg, ``Temperature effects on
adhesive bond strengths and modulus for commonly used spacecraft structural adhesives,'' Jet Propulsion
Laboratory (Pasadena, CA, U.S.A.) technical report.

\bibitem{Hashimoto80}T. Hashimoto and A. Ikushima, {\it Rev. Sci. Inst.} {\bf 51}, 378 (1980).

\bibitem{stycastDatasheet}Emerson and Cumings Stycast{\textsuperscript{\textregistered}} 2850FT data sheet.





\end{thebibliography}
\end{document}